\documentclass[journal, 10pt]{IEEEtran}
\usepackage{graphicx}
\usepackage{amsmath}
\usepackage{amssymb}
\usepackage{subfigure}
\usepackage[noadjust]{cite}
\usepackage[warn]{textcomp}
\usepackage{multirow}
\usepackage{booktabs}
\usepackage{tabularx}
\usepackage{color,soul}
\usepackage[detect-all=true]{siunitx}
\usepackage{verbatim}
\usepackage{floatrow}
\usepackage{caption}
\usepackage{float}
\usepackage{siunitx}
\restylefloat{table}
\usepackage{amssymb}
\usepackage[tableposition=top]{caption}
\usepackage{dblfloatfix}
\usepackage{float}
\floatstyle{plaintop}
\restylefloat{table}
\usepackage{booktabs}
\usepackage{multirow} 
\usepackage{multicol}
\usepackage{rotating}
\usepackage{amsmath}
\usepackage[font=small,labelfont=bf]{caption}

\usepackage{lettrine}
\usepackage{lipsum}
\usepackage{setspace}
\newcommand{\ie}{\textit{i.e.}}
\newcommand{\eg}{\textit{e.g.}}

\newcommand{\re}{\textcolor{red}}

\begin{document}
\IEEEoverridecommandlockouts
{\setstretch{1}
\title{\huge{A Single-MOSFET MAC for Confidence and Resolution (CORE) Driven Machine Learning Classification \re{}}
   \vspace{0.15in}
}
\author{Farid Kenarangi,~\IEEEmembership{Student Member,~IEEE} and 
			Inna Partin-Vaisband,~\IEEEmembership{Member,~IEEE} \\
\thanks{The authors are with the Department of Electrical and Computer Engineering,
	University of Illinois at Chicago, Chicago, IL 60607 USA (e-mail:
	fkenar2@uic.edu; vaisband@uic.edu).}
\vspace{-0.3in}
}
\maketitle
\begin{abstract}
Mixed-signal machine-learning classification has recently been demonstrated as an efficient alternative for classification with power expensive digital circuits. In this paper, a high-COnfidence high-REsolution (CORE) mixed-signal classifier is proposed for classifying high-dimensional input data into multi-class output space with less power and area than state-of-the-art classifiers. A high-resolution multiplication is facilitated within a single-MOSFET by feeding the features and feature weights into, respectively, the body and gate inputs. High-resolution classifier that considers the confidence of the individual predictors is designed at \SI{45}{\nano\meter} technology node and operates at \SI{100}{\mega\hertz} in subthreshold region.
To evaluate the performance of the classifier, a reduced MNIST dataset is generated by downsampling the MNIST digit images from 28 $\times$ 28 features to 9 $\times$ 9 features. The system is simulated across a wide range of PVT variations, exhibiting nominal accuracy of 90\%, energy consumption of \SI{6.2}{\pico\joule} per classification (over 45 times lower than state-of-the-art classifiers), area of 2,1\SI{79}{\square\micro\meter} (over 7.3 times lower than state-of-the-art classifiers), and a stable response under PVT variations.

\smallskip
\emph{Index Terms}---machine learning hardware, mixed-signal classifiers, confidence-level, high resolution, high-dimensional data, multi-class classification, linear classifiers, logistic regression, subthreshold.
\end{abstract}
\section{INTRODUCTION}
\lettrine{L}{OW} power machine learning (ML) classifiers are playing an important role in enabling edge computation of ML algorithms. Wide variety of applications benefit from these compact classifiers, such as, Internet of Things (IoT) devices, wireless sensor networks (WSNs), smart home technologies, autonomous transportation, and security systems.

Existing on-chip classifiers can be categorized into two major domains: digital and mixed-signal \cite{sze2017designing}. A digital classifier is typically fed with binary inputs (\ie, features) and uses binary feature weights, all obtained by sampling and quantizing corresponding analog signals. The classification accuracy with digital classifiers increases with the increasing number of bits assigned for features and weights. These highly accurate digital classifiers however exhibit significant power consumption and physical size and are often not suitable for power limited applications, such as battery powered sensors and those other IoT devices that are wirelessly powered and powered from harvested energy. Alternatively, mixed signal classifiers aim to reduce the area and power consumption of the conventional digital classifiers by directly using the analog input data for classification \cite{wang2017low}. The inherent need for data conversion with power hungry analog-to-digital converters (ADCs) is therefore mitigated with mixed-signal classifiers \cite{wang2017low}.

Another concern in modern classifiers is the high dimensionality of data. Classifying data in a high-dimensional space often results in a prohibitively high data movement among memory and computing circuit components. This, in turn, significantly increases power consumption in both, mixed-signal and digital classifiers \cite{chen2017eyeriss,hameed2010understanding,horowitz20141,whatmough201728nm,bankman2018always,jeon201723,bang201714,desoli201714,bong201714,shin201714,moons201714,park20154,farid2018memory}. To reduce the data movement, several approaches for in-memory ML computation have recently been proposed. Recent state-of-the-art in-memory classifiers typically exhibit accuracy between $90\%$ and $96\%$ and energy dissipation in the range of \SI{210}{\pico\joule}/decision to \SI{879}{\pico\joule}/decision \cite{zhang2017memory,gonugondla201842pj,kang2018memory,tang2018scaling} for typical image recognition datasets. Emerging device technologies are also being considered for providing power and area efficient alternatives for the conventional CMOS based classifiers. Accuracy of 90\% and energy of \SI{25}{\pico\joule} per decision has been recently reported in \cite{querlioz2011simulation,kenarangi2019leveraging}. 
	
\begin{figure}[t]
	\centering
	\includegraphics[width=0.6\columnwidth]{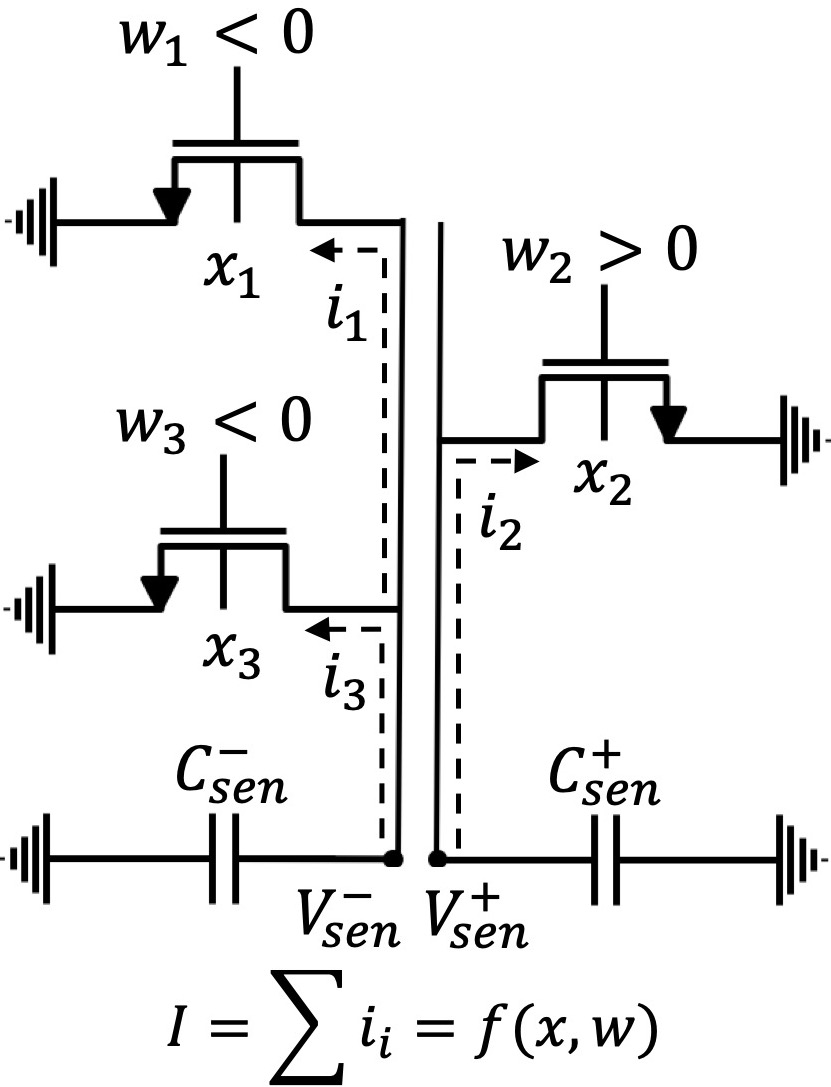}	
	\caption{Schematic of the proposed linear binary classifier. Features and feature weights are fed into, respectively, the body and gate input of the individual MOS transistors. Separate lines are used for positive and negative weights.\vspace{-10pt}}
	\label{fig:proposed_without}
\end{figure}

To enable high-resolution feature-weight multiplication, a theoretical framework that comprises circuits, models, and algorithms is proposed. To the best of the authors knowledge, this paper is the first to report a mixed-signal high-resolution classifier, utilizing MOSFET body terminals.
The schematic of the proposed configuration is shown in Fig. \ref{fig:proposed_without}. With this approach, body bias of the MOSFETs is controlled by the individual ML features, the gate inputs are fed by the absolute value of a corresponding feature weight, and the sign of the weights is taken into account by considering separate lines for the positive and negative feature and feature weight product. 
A $K$-class classification with $N$ features is therefore realized with $N$ rows and $K$ columns multiplication and accumulation (MAC) array, where each column serves as an independent binary classifier. These individual binary classifiers are combined using one-versus-all technique \cite{rifkin2004defense}, requiring $\frac{1}{2}(K-1)$ times less binary classifiers than the state-of-the-art classifier in \cite{zhang2017memory} and orders of magnitude less transistors, as compared with other existing mixed-signal classifiers \cite{gonugondla201842pj,tang2018scaling}.

Another primary contribution is in decision making domain. With analog classifiers, binary predictions are typically made based on the relative magnitude of signals between the positive and negative sensing lines. With this traditional approach, the sensing line with the highest voltage drop is assumed to exhibit the correct classification and the confidence level of the decision is not considered. Alternatively, a small difference in these voltage drop values indicates low confidence level of the individual binary predictions and thus, a higher probability of an erroneous final decision. To address this ML integrity issue, a confidence driven classification is proposed in this paper. With this approach, the difference among the magnitudes of the sensing line voltage drops is considered for capturing the confidence of the individual predictions. 

Finally, the proposed system is designed in subthreshold region, exhibiting a power efficient alternative for the traditional classifiers. The classifier is demonstrated on Modified National Institute of Standards and Technology (MNIST) dataset \cite{lecun1998mnist} of 10-class digit images. Based on SPICE circuit level simulation results, MNIST data is classified with 90\% accuracy using $81\times10$ transistors and exhibits power consumption of only \SI{6.2}{\pico\joule} per decision.

The rest of the paper is organized as follows.
In Section \ref{sec:binary_classifier}, the proposed high-resolution \textit{binary} classifier and linearization technique are described. Fabrication considerations are also discussed in this section. Based on the proposed binary classifier, a \textit{multi-class} high-resolution classifier is designed and demonstrated with MNIST dataset, as described in Section \ref{sec:multi-class}. Confidence driven classification is also explained in Section \ref{sec:multi-class}.
Circuit design and simulation results of the multi-class high-COnfidence high-REsolution (CORE) classifier using one-versus-all technique are presented in Section \ref{sec:sim_results}. The paper is summarized in Section \ref{sec:summary}.
\section{THE PROPOSED LINEAR BINARY CLASSIFIER} \label{sec:binary_classifier}
In this section, the proposed linear binary classifier is described. The software level design framework is provided in Section \ref{sec:ML_sec}. The circuit, and fabrication level considerations are presented in, respectively, Sections \ref{sec:circuit} and \ref{sec:fabrication}.
\subsection{Design Framework}\label{sec:ML_sec}
Reliability, power consumption, and physical size of on-chip classifiers are all primary concerns in modern ML ICs. The proposed framework is designed to meet accuracy specifications of modern classification problems in a cost effective manner. 
Linear algorithms are exploited in this paper for training a supervised binary classifier, optimizing the system for linearly separable input data.

With a multivariate linear classifier, the system response $Z$ is a linear combination of $N$ input features $x=({x_1}, {x_2}, ...,{x_N})$ and model weights $w=({w_1},
{w_2}, ...,{w_N})$,
\begin{equation} \label{eq:Z}
Z=\sum_{i=1}^{N}w_{i}\cdot x_{i}, \quad Z\in \mathbb R.
\end{equation}
The model weights are determined during supervised training by minimizing the prediction error between the system response, $Z$, and a corresponding true value in the labeled training dataset. Logistic regression (LR) -- a common supervised linear ML model -- is used for training the proposed classifier based on gradient descent algorithm \cite{nelder2004generalized}. LR is preferred due to its simple implementation and superior performance on MNIST dataset as compared with other classifiers. In inference, a probability threshold of $0.5$ is used for predicting system response to input data, exhibiting a simple on-chip implementation, 
\begin{equation} \label{eq:decision}
\resizebox{.91\hsize}{!}{$\hat{y}=\text{sign}(Z)=\text{sign}(\sum_{i=1}^{N}w_{i} \cdot x_{i})
	=\left\{
	\begin{array}{ll}
	1,  \quad  &\:Z\ge0\\
	-1, \quad  &\:Z <0.\\
	\end{array}
	\right.$}
\end{equation}
The described logistic regressor with the probability threshold of 0.5 is referred to as logistic classifier. 
The accuracy of the proposed logistic classifier is evaluated as a percentage of all the correct predictions out of the total number of test predictions. The proposed ML flow and the preprocessing steps are explained below.

\subsubsection{Dataset} MNIST database is a large image dataset, commonly used for evaluating the effectiveness of ML hardware. MNIST contains images of 70,000 handwritten digits, ranging between 0 to 9. Each digit comprises 784 ($28 \times 28$) image pixels. 
The training and test datasets comprise, respectively, 60,000 and 10,000 digits. Out of the 60,000 training observations, 45,000 and 15,000 digits are used for, respectively, training and validating the proposed system.
\begin{figure}[t]
	\centering
	\includegraphics[width=0.95\columnwidth]{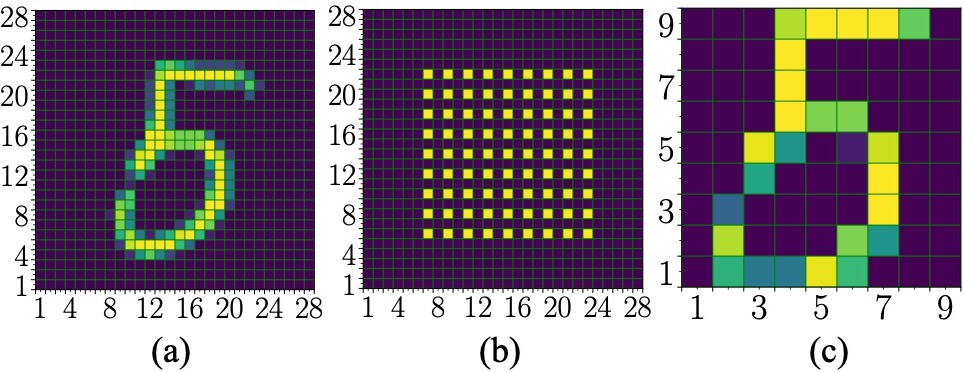} 
	\vspace{-5pt}
	\caption{Handwritten digits from MNIST dataset, (a) original image (default resolution with $28\times 28$ features), (b) selected features shown with light squares, and (c) downsampled image with $9 \times 9$ features.\vspace{-20pt}}
	\label{fig:down_sample_figure}
\end{figure}
\subsubsection{Feature selection and  downsampling} Each image pixel of the individual digits in the training set is considered as an ML feature and used for training the classifier. 
To reduce the power and area overheads, those redundant features that are not essential for digit classification are eliminated.
To determine the preferred number of observed features, the dataset is downsampled to $N\leq784$ features ($N=6^2,\space 7^2,\dots.\space 28^2$) and classification accuracy is obtained for the downsampled data in Python. 
To efficiently classify MNIST digits with the proposed classifier, $N=81$ $(9\times 9)$ is preferred, corresponding to 90\% accuracy.
The original and downsampled ($N=81$) digit images are shown in Fig. \ref{fig:down_sample_figure}.
\subsection{Circuit Level Considerations} \label{sec:circuit}
The primary goal in a linear binary classification is to accurately and efficiently perform the dot product operation of the features and feature weights, as described in \eqref{eq:decision}. To simplify the circuit level design, the result shown in \eqref{eq:decision} is formulated as the signed addition of positive, $Z^+$, and negative, $Z^-$, feature-weight products,
\begin{equation} \label{eq:target}
\resizebox{.9\hsize}{!}{$\hat{y}=\text{sign}(\overbrace{ \sum_{w_i>0}x_i \cdot w_i}^{Z^+} + \overbrace{\sum_{w_j<0}x_j \cdot w_j}^{Z^-})
=\left\{\begin{array}{ll}
1,\:Z^+\ge Z^-\\
-1,\:Z^+ < Z^-.\\
\end{array}\right.$}
\end{equation}

The individual positive and negative feature-weight products are accumulated within the positive, $V^+_{sen}$ and negative $V^-_{sen}$ sensing lines, yielding the basic ML multiplication and accumulation (MAC) operation, as shown in Fig. \ref{fig:proposed_without}. For each feature-weight multiplication, a single-MOSFET is connected to either the positive or negative sensing line, as determined by the sign of the corresponding feature weight. For example, for $w_1<0$, the corresponding multiplier MOSFET is connected to the negative sensing line. 

To capture the voltage drops across the sensing lines, sensing capacitors (\ie, $C_{sen}$) are connected to the individual sensing lines. The size of a sensing capacitor is determined proportionally to sensing line current, increasing with the number of transistors (\ie, features) connected to the line. To classify tasks with higher number of features, larger sensing capacitors are therefore required, limiting the scalability of the system. To provide a power efficient and scalable solution, the transistors are biased in near/subthreshold operation region, significantly limiting the current through the sensing lines. In the case of MNIST classification with $9 \times 9$ features, capacitance of only \SI{50}{\femto\farad} is utilized per sensing line. 
\begin{figure}[t]
	\centering
	\includegraphics[width=1\columnwidth]{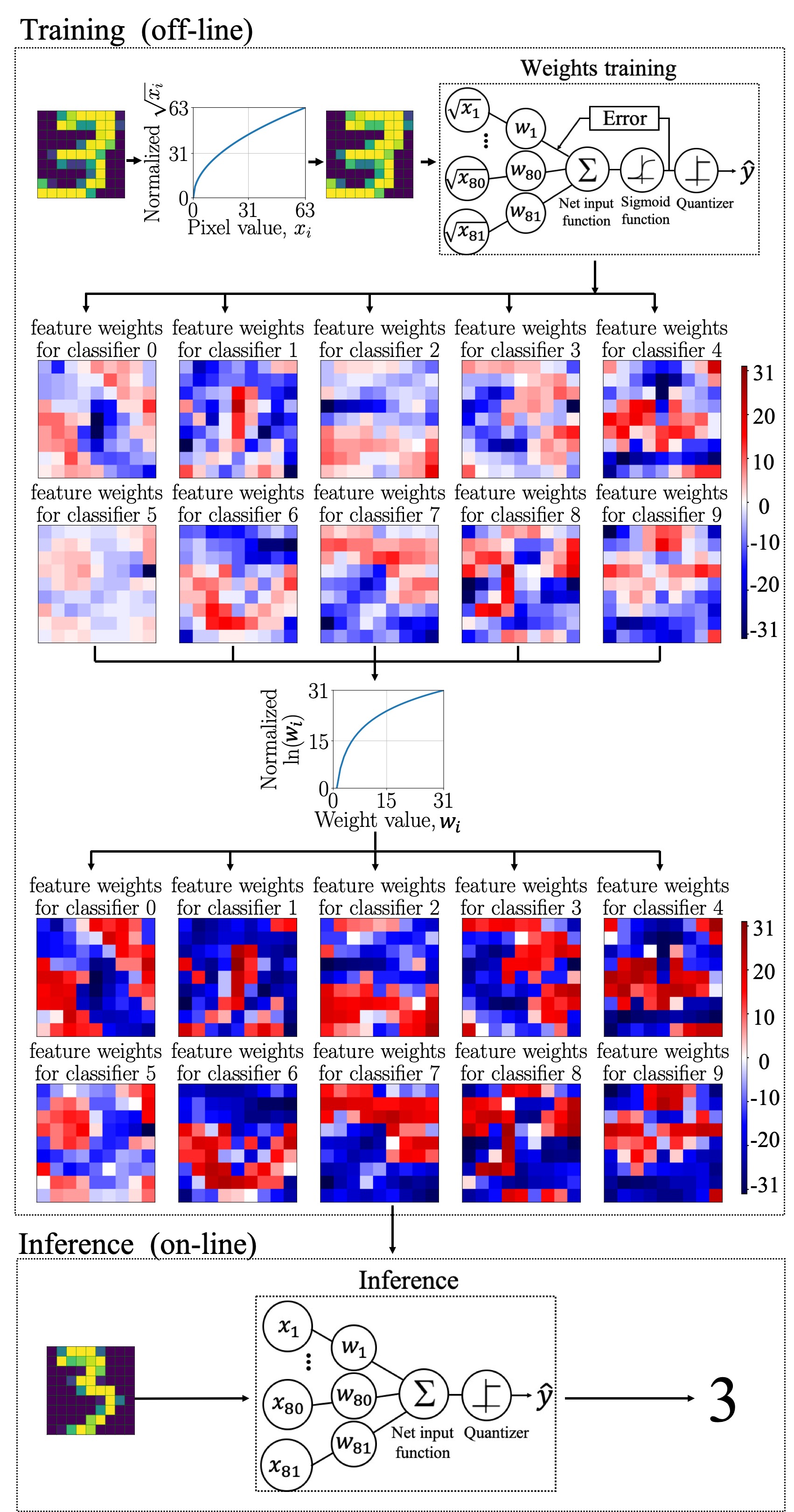} 
	\vspace{5pt}
	\caption{Linearization flow. To account for the non-linear dependence of the drain current on the body bias, $I_{sub} \propto \sqrt{V_{bs}}$, the model is trained to make predictions based on the square root values of the original features. The optimized weights are logarithmically adjusted, mitigating the exponential dependence of the subthreshold current on the gate bias.\vspace{-10pt}}
	\label{fig:flow_linear}
\end{figure}

A primary concern with near/subthreshold operation is the exponential dependence of the drain current on the body and gate biases \cite{gray2001analysis}, 
\begin{equation} \label{equ:sub-threshold}
I_{sub}=\frac{W}{L}I_t\, \text{exp}(\frac{V_{gs}-V_{th}}{nV_T})[1-\text{exp}(-\frac{V_{DS}}{V_T})],
\end{equation}
where $I_t$ is the sub-threshold current at $V_{gs}=V_{th}$, $n$ is the sub-threshold slope, and $V_T$ is the thermal voltage. Note that body voltage dependence is embedded in the threshold voltage, $V_{th}\propto  \sqrt{V_{bs}}$. Considering that $V_{DS}>>V_T$, the expression in \eqref{equ:sub-threshold} can be simplified as,
\begin{equation} \label{equ:sub-threshold_reduced}
I_{sub}=\frac{W}{L}I_t\, \text{exp}(\frac{V_{gs}-V_{th}}{nV_T}), V_{DS} >> V_T.
\end{equation} 

To mitigate the non-linear dependence of the drain current on the weight-feature dot product (see \eqref{eq:Z}), a novel training flow is proposed (see Fig. \ref{fig:flow_linear}). To account for the non-linear dependence of the drain current on the bias voltage, the model is trained with square root values of the default features ($x_i \rightarrow \sqrt{x_i})$). Thus, the extracted feature weights, $w$, are optimized for classifying the MNIST dataset transformed into half-order polynomial space.  
Alternatively, to account for the non-linear dependence of the drain current on the gate voltage, the feature weights are logarithmically adjusted ($w_i \to \ln(w_i)$), exhibiting $V_{gs}\propto \ln(w)$. Using the first-order approximation of $\exp(\sqrt{V_{bs}})\approx 1 + \sqrt{V_{bs}}$, the current in this case is expressed as,
\begin{equation} \label{log_map}
I_{sub} \propto \exp(\ln(V_{gs})) \cdot \exp(\sqrt{V_{bs}})\propto V_{gs} \cdot \sqrt{V_{bs}} \propto w \cdot \sqrt{x}.
\end{equation}
In inference, the current model is exploited for making prediction based on the square root values of the original features, as trained offline, yielding 90\% accuracy across the MNIST test set, as detailed in the following sections. 
\subsection{Fabrication Costs} \label{sec:fabrication}
In the proposed linear binary classifier, the body and gate terminals are fed by, respectively, the input features and corresponding feature weights. Each multiplication is, therefore, executed by a single-MOSFET, significantly reducing the power and area costs (despite the overhead of the triple-well technology) and complexity (as determined by number of transistors) of the classifier in comparison to the existing state-of-the-art mixed-signal classifiers \cite{wang2017low,zhang2017memory}.

Conventional twin-well fabrication process is illustrated in Fig. \ref{fig:fab}(a). This process is designed to provide a single voltage connection to all the n-type and p-type body terminals. Alternatively, to connect body terminals of the individual multiplier transistors to different voltage levels, a specialized fabrication process is required. One way to individually bias numerous body terminals, is by fabricating with triple-well process (see Fig. \ref{fig:fab}(b)), which is commonly used in high-performance, low-power ICs \cite{tschanz2002adaptive,taco2016low} and for reducing substrate noise in mixed-signal circuits \cite{to2001comprehensive}. In a p-substrate triple-well process, an additional deep n-well is used to isolate the p-well of each MOSFET from the p-substrate, allowing an independent body terminal connection for each MOSFET.
The triple-well structure has been demonstrated to provide better noise characteristics as compared with the traditional twin-well structure, without increasing the gate leakage \cite{ogasahara2013supply}. Alternatively, the triple-well structure, exhibits additional fabrication costs and area overheads that needs to be considered. Layout of a four-transistor block in twin-well and triple-well process is shown in, respectively, Fig. \ref{fig:fab}(c) and Fig. \ref{fig:fab}(d). With the triple-well configuration, the area is increased 3 times as compared with the twin-well process in \SI{45}{\nano\meter} CMOS technology.
\begin{figure}[t]
	\centering
	\includegraphics[width=0.9\columnwidth]{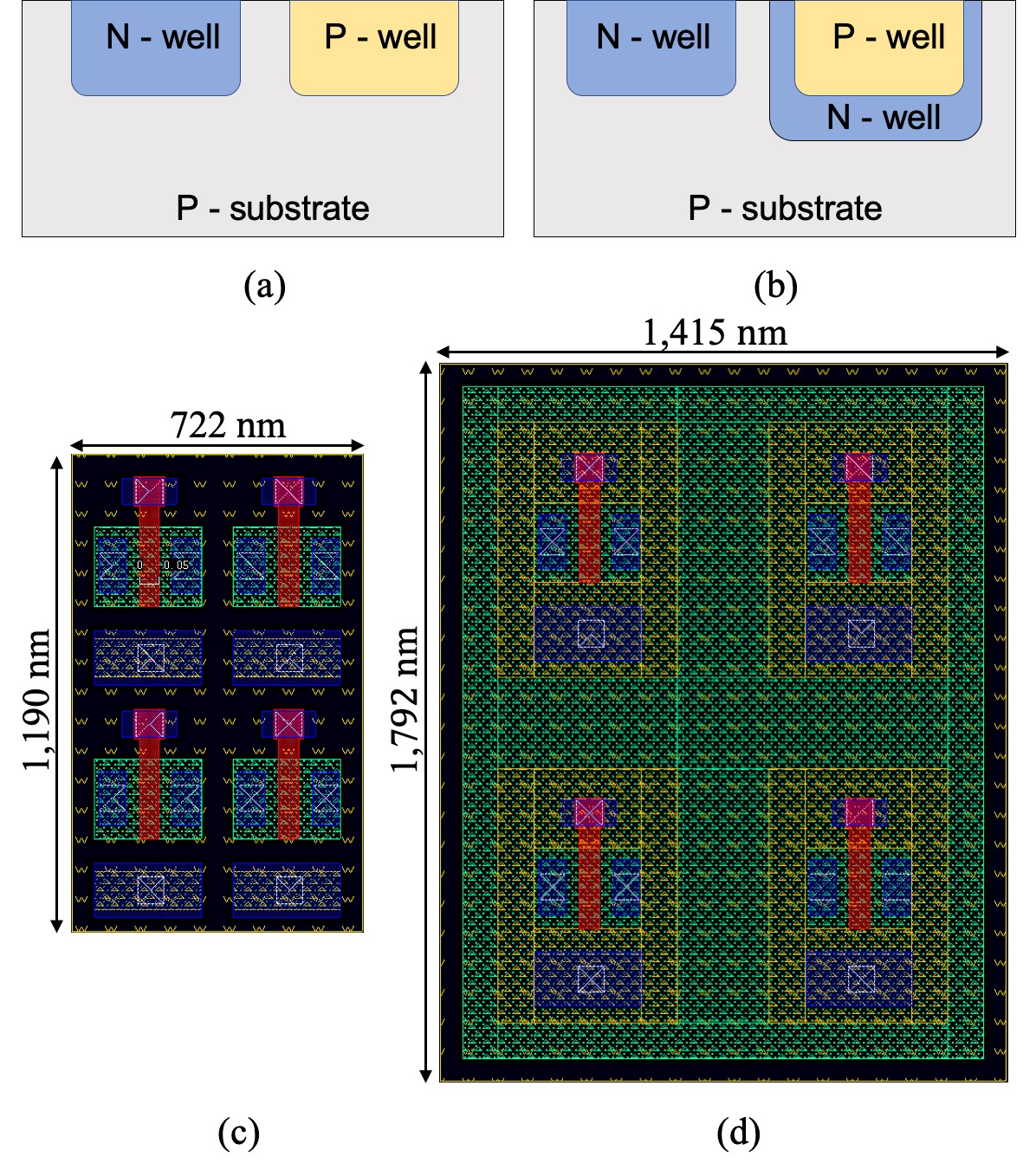}
	\caption{Two commonly used fabrication processes, (a) twin-well process, and (b) triple-well process with p-substrate. Layout of four transistors with W = \SI{200}{\nano\meter} and L = \SI{50}{\nano\meter} in the (c) twin-well, and (d) triple-well configurations.}	
	\label{fig:fab}
\end{figure}
\section{CORE BASED MULTI-CLASS CLASSIFIER} \label{sec:multi-class}
A multi-class classifier is designed based on multiple linear binary classifiers, as presented in Section \ref{sec:binary_classifier}. 
Confidence driven approach that addresses the integrity of multi-class classification is described in Section III.A.
The transistor level implementation of the proposed CORE classifier is presented in Section III.B.

\subsection{Confidence Driven Classification: OVO versus OVA} \label{sec:decision}
 Two typical approaches for designing a multi-class classifier based on multiple binary classifiers are one-versus-one (OVO) and one-versus-all (OVA) \cite{aly2005survey}. With OVO approach, all pairwise combinations of the output classes are evaluated with the individual binary classifiers. 
 Thus, a $K$-class classification with OVO approach requires $\frac{1}{2}K(K-1)$ binary classifiers, increasing the system complexity and power and area costs quadratically with the number of classes. For example, for classifying 10 digits with OVO approach, 45 $i$-verus-$j$, ($i,j\in \{0,1, 2, .., 9\}$) binary classifiers are required.
 Alternatively, in case of a 1,000 class classification, about half a million of binary classifiers are required. The final decision with OVO technique is extracted using majority voting approach \cite{levin1995introduction}: each binary classifier votes independently for a certain class and the final decision is made based on the class with highest number of votes. 
 
 Alternatively, with OVA approach, each binary classifier discriminates between a single class and the rest of the classes. The required number of binary classifiers with OVA increases linearly with the number of classes, facilitating a more power and area efficient classification of high-dimensional data.
 In addition, a probability score of correct discrimination is inherently provided with OVA and can be extracted for the individual binary classifiers. Thus, a $K$-class OVA classifier can be designed with as few as $K$ binary classifiers, seamlessly accounting for the confidence level of the individual predictions. The confidence driven decision, $d$, with the OVA classifier is determined as, 
\begin{equation} \label{eq:final_decision}
d=i|p_i=\text{max}(p_i),\space i=\space 0,\space 1,\space...,\space K-1,
\end{equation}
where $p_i$ is the confidence level of the $i^{th}$ binary classifier.

From circuit level perspective, the confidence level with OVA approach can be determined as the difference of the voltage drops across the positive and negative sensing lines, \eg, $p_i \propto \Delta V^+_{sen}(i) - \Delta V^-_{sen}(i) \triangleq \Delta V_{sen}(i)$.
\begin{equation} \label{eq:refinal_decision}
d=i|(\Delta V_{sen}(i))=
\max_{j=0}^{K-1} \left\{\ (\Delta V_{sen}(j)\right\}.
\end{equation}
\begin{figure}[t]	
	\centering
	\includegraphics[width=1\columnwidth]{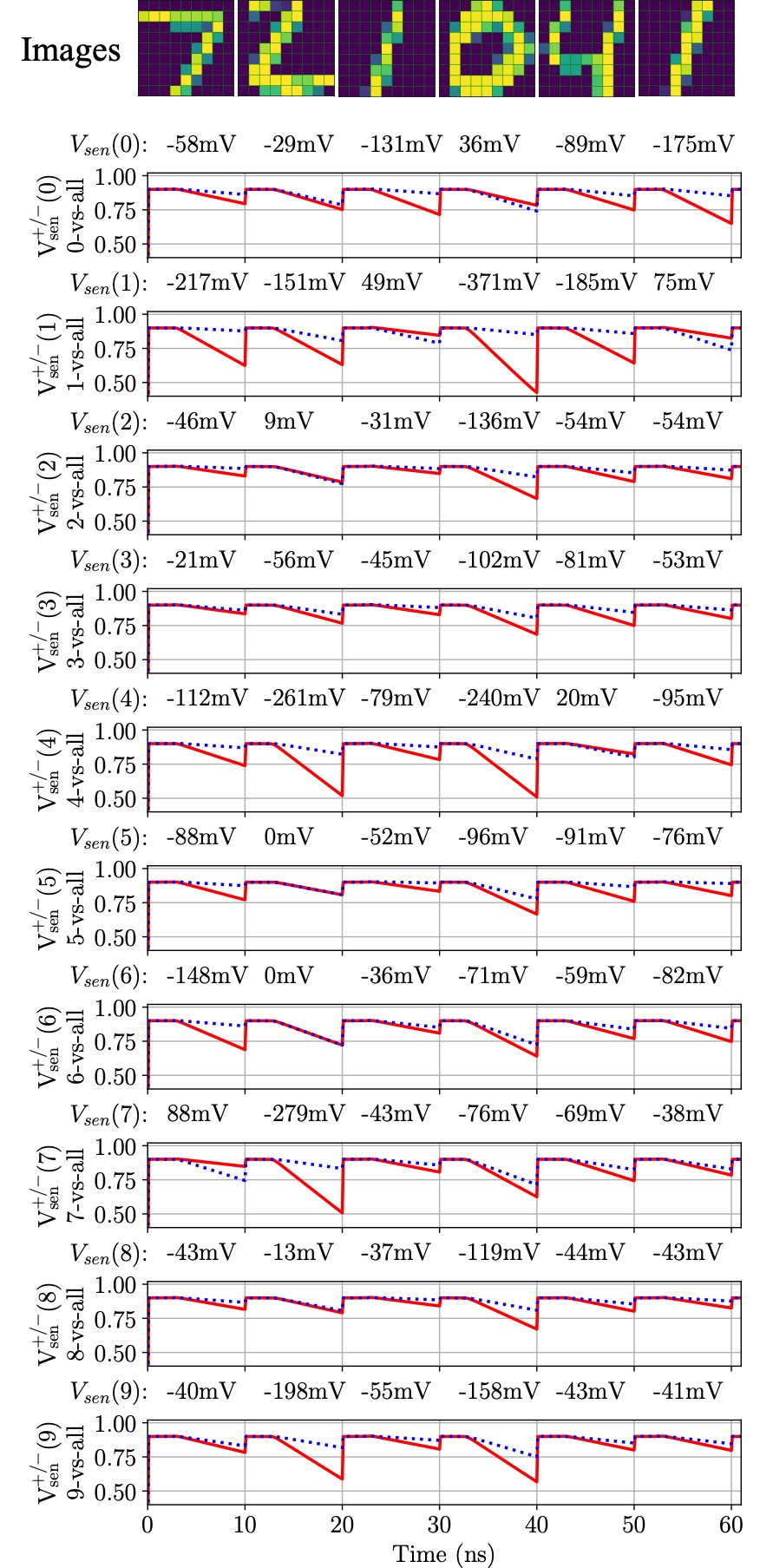} 
	\caption{Voltage waveforms of the sensing lines during the precharge (\ie, ($0-\SI{2.5}{\nano\second}$), 
	($10-\SI{12.5}{\nano\second}$), 
	($20-\SI{22.5}{\nano\second}$), 
	($30-\SI{32.5}{\nano\second}$), 
	($40-\SI{42.5}{\nano\second}$), 
	and 
	($50-\SI{52.5}{\nano\second}$)
	 ) and classification stages (\ie, 
	 ($\SI{2.5}{\nano\second}-\SI{10}{\nano\second}$),
	 ($\SI{12.5}{\nano\second}-\SI{20}{\nano\second}$),
	  ($\SI{22.5}{\nano\second}-\SI{30}{\nano\second}$),
	  	 ($\SI{32.5}{\nano\second}-\SI{40}{\nano\second}$),
	  		($\SI{42.5}{\nano\second}-\SI{50}{\nano\second}$),
	   and ($\SI{52.5}{\nano\second}-\SI{60}{\nano\second}$)
	    for six consecutive input features (\ie, 7, 2, 1, 0, 4, and 1).} 
	\label{fig:waveforms}
\end{figure}
To extract the class with highest confidence level, a light-weight comparator is designed, as described in the next subsection, yielding a power and area efficient alternative for the traditional voter circuits.

\subsection{Circuit Level Design and Simulation Results}
The proposed multi-class classifier is designed in SPICE and demonstrated based on the reduced MNIST dataset.  
The OVA circuits and architecture of the MOSFET array are described in this section.

\subsubsection{MAC Array} To classify the downsampled $9\times 9$ MNIST digits, ten binary classifiers (see Section \ref{sec:binary_classifier}) are co-designed in SPICE, yielding a $81 \times 10$ MOSFET array.
Each of the transistors within the MAC array is exploited for generating a single feature-weight product. During inference, the $V^+_{sen}$ and $V^-_{sen}$ lines are precharged to $V_{DD}$ prior to each prediction. All the input features and feature weights are connected simultaneously to, respectively, the body and gate terminals of the individual multiplier transistors, facilitating a parallel classification process within all the ten binary classifiers. As a result, 20 different voltage drop values (\ie, $V^+_{sen}(i)$, $V^-_{sen}(i)$, $i=0,\space 1,\dots,\space 9$) are generated on the individual sensing lines, as shown in Fig. \ref{fig:waveforms}. The voltage waveforms of the positive and negative sensing lines are illustrated by, respectively, the blue dotted and solid red lines. 
These voltage drops are exploited by the confidence level extractor, generating the final classifier decision. The confidence level of each binary classifier (as determined based on \eqref{eq:refinal_decision}) is shown in Fig. \ref{fig:waveforms}, as noted on top of each plot. For example, for digit 7, the $7^{th}$ (7-vs-all) binary classifier has as expected the highest confidence level. The final decision is generated by the confidence level extractor.
\begin{figure}[t]	
	\centering
	\includegraphics[width=1\columnwidth]{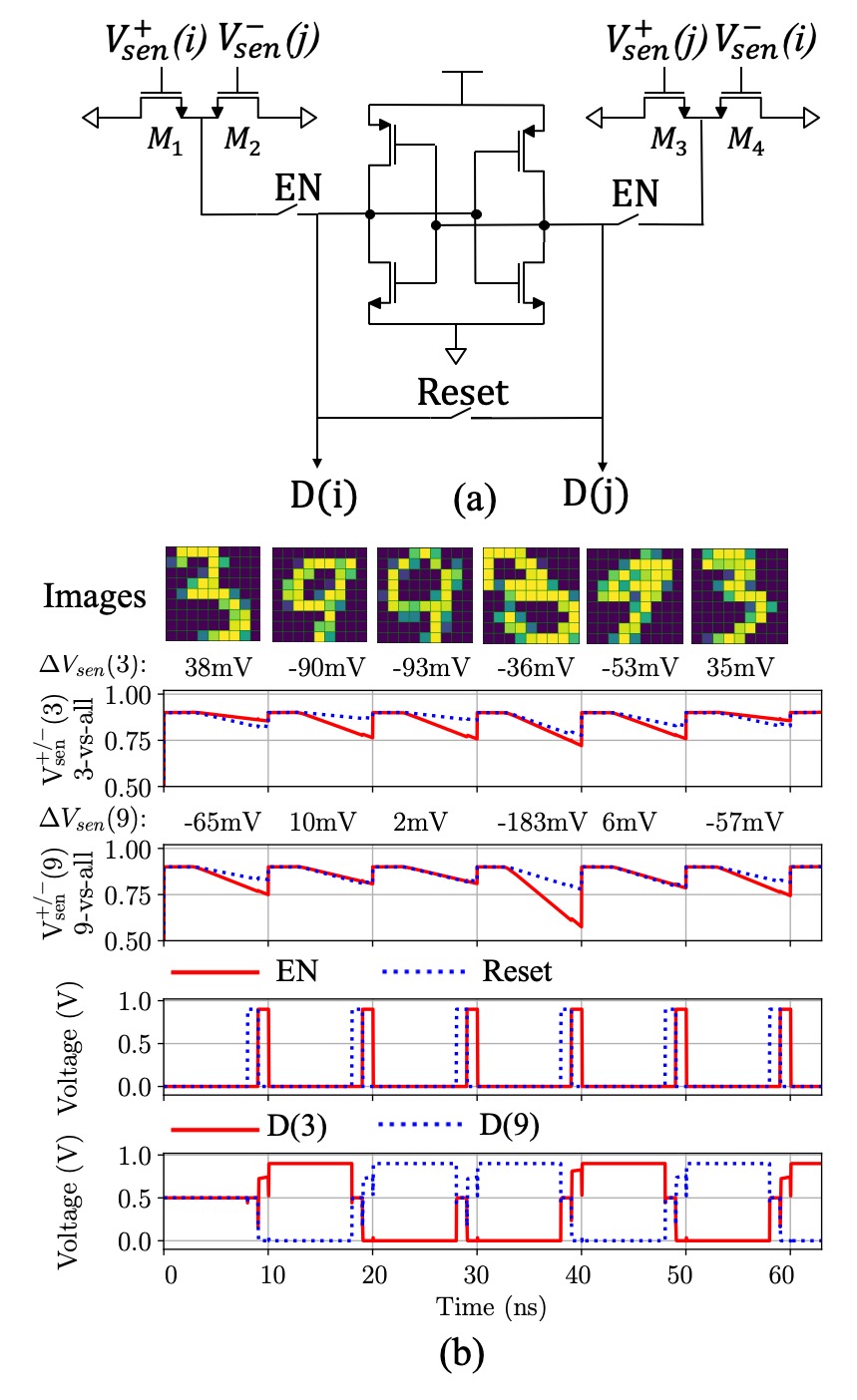} 
	\caption{Schematic of the proposed confidence level extractor. The confidence level of the $i^{th}$ and  $j^{th}$ binary classifiers is extracted and the prediction of the most confident classifier is used as the final decision.	\vspace{-10pt}} 
	\vspace{-5pt}
	\label{fig:MOSFET_confidence}
\end{figure}
\begin{figure*}[t]	
	\centering
	\includegraphics[width=1\columnwidth\vspace{-5pt}]{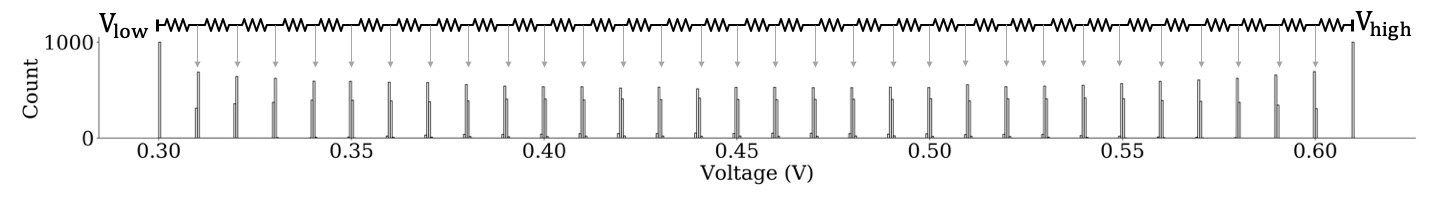} 
	
	\caption{Schematic of the resistive voltage divider. No sensitivity to process variations is observed with 1K-run Monte-Carlo simulation.\vspace{-15pt}} 
	\label{fig:res_divider}
\end{figure*}
\begin{figure}[t]	
	\centering
	\includegraphics[width=1\columnwidth]{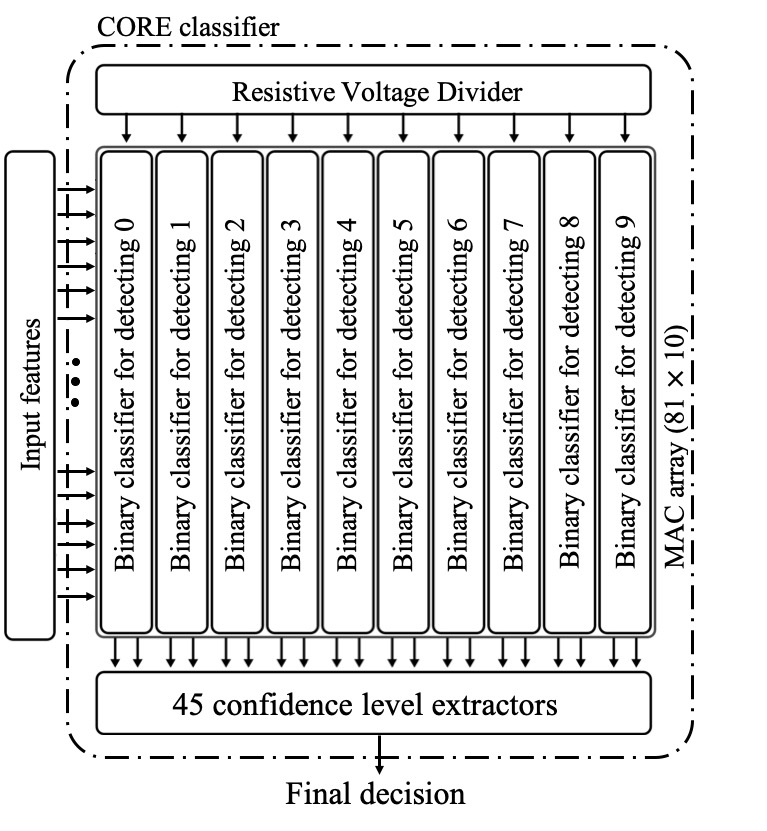} 
	\vspace{-10pt}
	\caption{Schematic of the proposed classifier, comprising voltage divider, MOSFET array, and confidence level extractors.\vspace{-15pt}} 
	\label{fig:overall}
\end{figure}
\subsubsection{Confidence Level Extractor} 
The schematic of a single confidence driven selector for a multi-class classification is presented in Fig. \ref{fig:MOSFET_confidence}(a). 
For a $K$-class classification, $\frac{1}{2} K(K-1)$ confidence driven selectors are required, yielding a total of 45 selectors for the MNIST dataset ($K=10$). 
The circuit is designed to compare the confidence levels of the binary classifiers ($i^{th}$ and $j^{th}$ classifier) and determine the classifier with higher confidence level,  
\begin{equation} \label{eq:conf_circuit}
\Delta V^{+}_{sen}(i)- \Delta V^{-}_{sen}(i)  <> \Delta V^{+}_{sen}(j)- \Delta V^{-}_{sen}(j).
\end{equation}
To simplify the circuit level implementation the subtraction in \eqref{eq:conf_circuit} is replaced with summation,  
\begin{equation}\label{eq:conf_circuit_sum}
\Delta V^{+}_{sen}(i) + \Delta V^{-}_{sen}(j) <> \Delta V^{+}_{sen}(j) + \Delta V^{-}_{sen}(i).
\end{equation}
Each summation in \eqref{eq:conf_circuit_sum} is captured with two parallel NMOS transistors, as shown in Fig. \ref{fig:MOSFET_confidence}(a).
To capture the result of $\Delta V^{+}_{sen}(i) + \Delta V^{-}_{sen}(j)$, the gate terminals of the transistors $M_1$ and $M_2$ are connected to, respectively, the positive sensing line of the $i^{th}$ classifier, $V^{+}_{sen}(i)$, and negative sensing line of the $j^{th}$ classifier, $V^{-}_{sen}(j)$. 
As a result, the drain current at $M_1$ and $M_2$ is proportional to $\Delta V^{+}_{sen}(i) + \Delta V^{-}_{sen}(j)$. 
Similarly, the gate terminals of the transistors $M_3$ and $M_4$ are connected to, respectively, $V^{+}_{sen}(j)$ and $V^{-}_{sen}(i)$, generating a drain current proportional to $\Delta V^{+}_{sen}(j) + \Delta V^{-}_{sen}(i)$.

To determine which side exhibits higher confidence level (\ie, sinks lower current), two back to back inverters are utilized. 
Voltage waveforms of the sensing lines, EN signal which enables the back-to-back inverters, Reset signal which resets the voltages stored on both sides of the back-to-back inverters and output signals are illustrated in Fig. \ref{fig:MOSFET_confidence}(b) for six consecutive classification periods. During the second, third, and fifth classifications, the left side of the inverters sink higher current than the right side. 
The left and right sides of the confidence selector, are, therefore, forced to, respectively, the low (\ie, $D(i)$ = 0) and high (\ie, $D(j)$ = 1) output voltage. 
Alternatively, during the first, fourth, and sixth classifications, the right side sinks higher current than the left side. Thus, left and right sides are forced to, respectively, the high (\ie, $D(i)$ = 1) and low ($D(j)$ = 0) output voltage. 

The correct functionality of the confidence level extractor depends on its symmetric structure and is highly sensitive to process variations. To mitigate process variations, larger pull-down ($W=5W_{min}$), pull-up ($W=15W_{min}$), and confidence extractor ($W=5W_{min}$) transistors are utilized for the confidence level extractor. With upsized transistors, the average accuracy degradation of the classifier is limited to 2\% under process variations, as described in the next section. Note that conventional methods such as extracting final classification results with $K$ analog-to-digital (ADCs) can also be leveraged, trading-off power efficiency for scalability (\ie, $\frac{1}{2}(K-1)$ times less confidence extractors).  

\subsubsection{Resistive Voltage Divider} 
The trained, quantized feature weights are generated using a resistive voltage divider (see Fig. \ref{fig:res_divider}).
In this configuration, the preferred voltage range ($V_{low}$, $V_{high}$) is divided into $2^n$ equal steps, where $n$ is the preferred quantization resolution. The gate bias range (\ie, ($\SI{300}{\milli\volt}, \SI{610}{\milli\volt}$) is quantized with 5-bit resolution into 31 equal steps of $\SI{10}{\milli\volt}$, as illustrated in Fig. \ref{fig:res_divider}. 
Poly resistors with sheet resistance of 7.8 $\Omega/\square$ are utilized.
The resilience of the voltage divider to process variations is evaluated with 1K-run Monte-Carlo simulation. Based on the simulation results, the circuit is highly resilient to process variations, exhibiting the average deviation of \SI{0.2}{\milli\volt} from the nominal wight values. The results of the 1K-run Monte-Carlo simulation are also shown in Fig. \ref{fig:res_divider}.
\begin{figure}[t]	
	\centering
	\hspace{-10pt}
	\subfigure[]{
	\includegraphics[width=0.48\columnwidth]{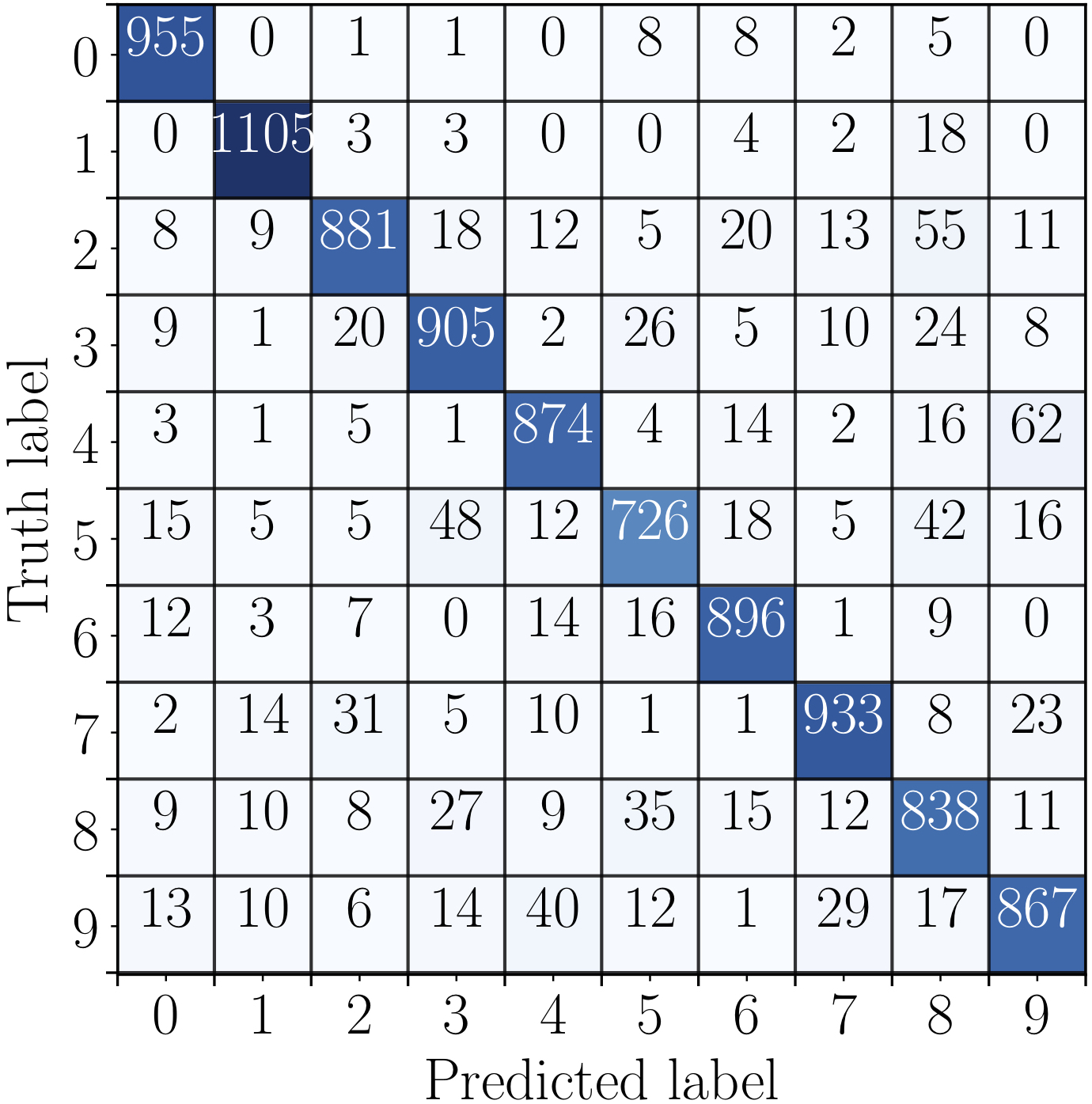} 
	\hspace{-10pt}
}
	\subfigure[]{
\includegraphics[width=0.48\columnwidth]{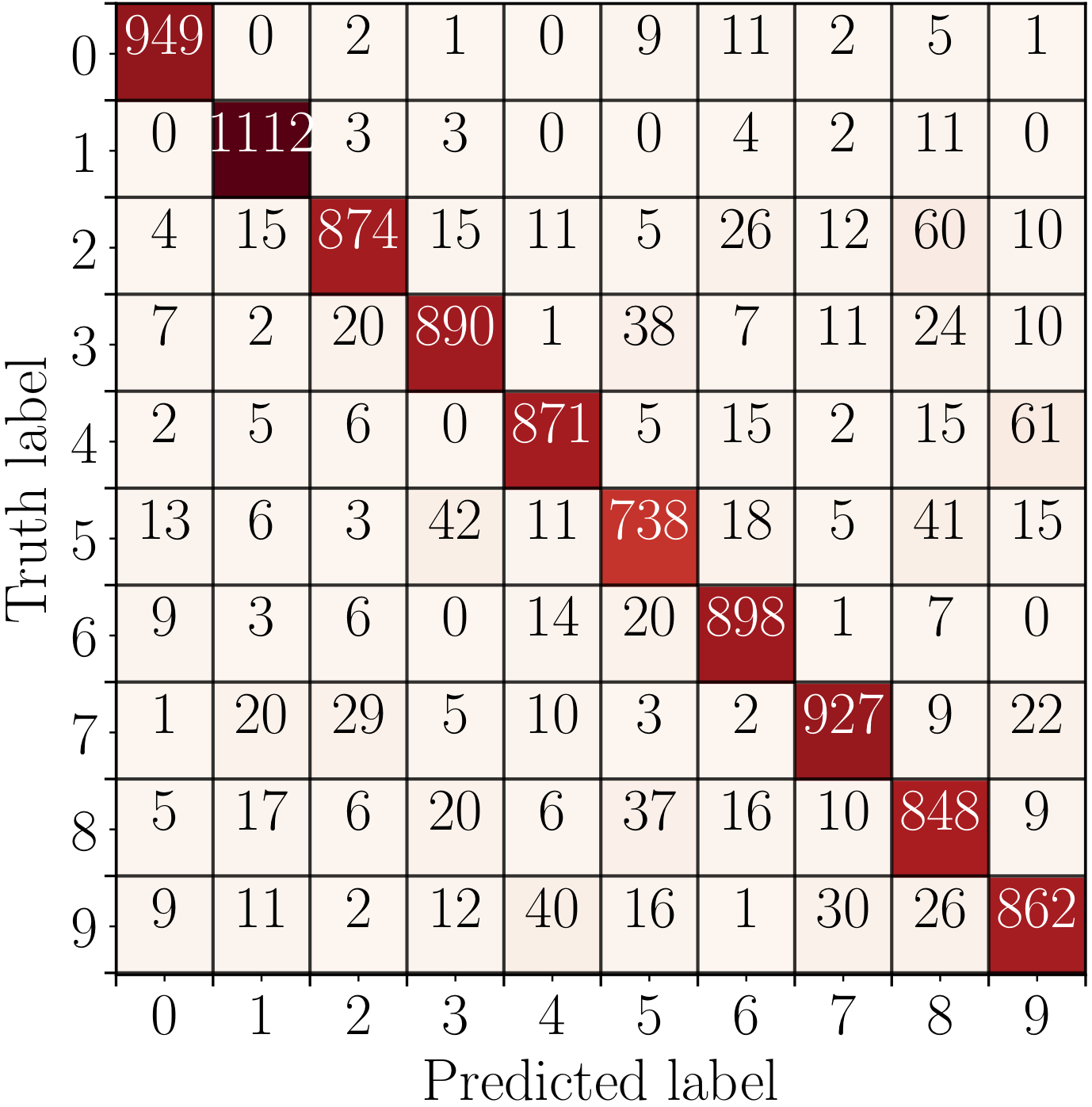} 
}
	\caption{Confusion matrix of the MNIST classification obtained in (a) Python (90\% accuracy), and (b) SPICE (90\% accuracy), exhibiting no accuracy degradation in SPICE as compared with Python results.} 
	\label{fig:conf_matrix}
\end{figure}

With this topology, no memory and data conversion units are required for storing and quantizing the weights. Alternatively, the weights provided with the voltage divider are not reconfigurable. Using a 32-to-1 multiplexer and a memory unit (\eg, SRAM), the circuit can be updated to provide reconfigurable feature weights \cite{lu201210}. The overall area of the classifier with reconfigurable weights is expected to increase by a factor of 4.2.   

\section{RESULTS} \label{sec:sim_results}
\subsection{System Characteristics}
A schematic of the integrated system is illustrated in Fig. \ref{fig:overall}, comprising voltage divider, MOSFET array, and confidence level extractors.
The proposed CORE classifier is designed at \SI{45}{\nano\meter} technology node with the nominal power supply voltage of \SI{0.9}{\volt} and threshold voltage of \SI{0.61}{\volt}. The body terminals of the MAC array are biased at voltage levels between \SI{0.2}{\volt} and \SI{0.8}{\volt}.
The area occupied by the classifier is 2,1\SI{79}{\square\micro\meter}, as estimated based on transistor count in SPICE. The MAC array comprises a total of $81\times10=810$ MOSFETs. The reduced, 81-feature MNIST dataset is classified with accuracy of 90\%  within a single \SI{10}{\nano\second} clock cycle of the system operation, exhibiting no accuracy degradation as compared with the validation accuracy in Python. 
The confusion matrices obtain based on Python and SPICE simulations are shown in, respectively, Fig. \ref{fig:conf_matrix}(a) and Fig. \ref{fig:conf_matrix}(b), exhibiting equal accuracy of 90\%.
The ML classifier generates predictions at \SI{100}{\mega\hertz} frequency, exhibiting an average energy consumption of \SI{6.2}{\pico\joule} per classification of a single digit. 
To maintain high prediction accuracy, 5 bits and 6 bits are assigned for quantizing, respectively, the feature weights and input features. By increasing the dimensionality of the proposed classifier, lower power and area overheads can be traded off for higher prediction accuracy, approaching the theoretical limit of 92\% for MNIST classification with linear ML algorithms and OVA decisioning scheme. The existing tradeoffs between dimensionality of the data and the accuracy, power, and area overheads are shown in Fig. \ref{fig:power_area}. The knee point of $N=81$ is selected in this paper to provide satisfactory accuracy results in an power and area efficient manner.
\begin{figure}[t]	
	\centering
	\includegraphics[width=1\columnwidth]{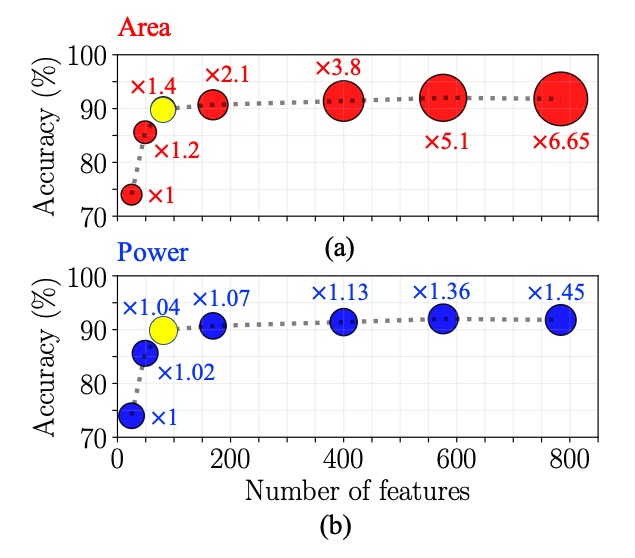} 
	\vspace{-10pt}
	\caption{Existing tradeoffs between the selected number of features, accuracy and (a) area, and (b) power overheads.}
	\label{fig:power_area}
\end{figure}
\subsection{Simulation Results}
Performance characteristics are listed in Table \ref{table:evaluation} for the proposed system along with the existing state-of-the-art mixed-signal classifiers \cite{wang2017low,zhang2017memory,gonugondla201842pj}.
Note the different dataset, MIT-CBCL, used in \cite{gonugondla201842pj}. Classification accuracy is a strong function of data. The accuracy comparison among \cite{gonugondla201842pj} and other classifiers in Table \ref{table:evaluation} is therefore less valuable, albeit the excellent accuracy demonstrated in \cite{gonugondla201842pj}.   
Alternatively, benefiting from the high-resolution multiplications and confidence driven predictions, CORE classifier exhibits significantly less transistor count, and thus lower power consumption and smaller IC area, as compared with the other state-of-the-art classifiers. 
For fair comparison, current time per decision and the system area normalized by squared form factor are also shown in Table \ref{table:evaluation}. The current time per decision with the proposed classifier is approximately 45 times lower as compared with other approaches. Similarly, area savings (albeit the triple-well technology) range between $\times$7.3 and $\times$89 as compared to other classifiers, as shown in Table \ref{table:evaluation}. Note that the operational frequency is scalable and can be adjusted based on the application needs and constrains.
{\renewcommand{\arraystretch}{1.5}
	\begin{table}[t]
		\resizebox{\textwidth}{!}{%
			\renewcommand{\thetable}{\Roman{table}}
			
			\caption{\textcolor{black}{System characteristics of the proposed and other existing mixed-signal ML classifiers.}}
			\vspace{-10pt}
			\label{table:evaluation}
			\centering
			\Huge	
			\begin{tabular}{|c|c|c|c|c|c|}				
				\hline	
				\multicolumn{2}{|c|}{} &\cite{wang2017low}&\cite{zhang2017memory}&\cite{gonugondla201842pj}&Current work\\ \hline
				\multicolumn{2}{|c|}{Dataset}& MNIST&MNIST&MIT-CBCL&MNIST\\ \hline
				\multicolumn{2}{|c|}{Technology}&\SI{130}{\nano\meter}&\SI{130}{\nano\meter}&\SI{65}{\nano\meter}&\SI{45}{\nano\meter}\\ \hline
				\multicolumn{2}{|c|}{Algorithm} & Ada-boost&Ada-boost&SVM&LR\\ \hline
				\multicolumn{2}{|c|}{Accuracy} & 90\%&90\%&96\%&90\%\\ \hline
				\multicolumn{2}{|c|}{Feature weight resolution} & 4b&1b&8b&5b \\ \hline
				\multicolumn{2}{|c|}{Feature resolution} & 5b&5b&8b&6b \\ \hline
				\multicolumn{2}{|c|}{Number of features} &48&81&256&81 \\ \hline
				\multicolumn{2}{|c|}{Supply voltage (V)} &1.2&1.2&0.675&0.9 \\ \hline
				\multicolumn{2}{|c|}{Speed (MHz)} &1.3&50&32&100 \\ \hline
				\multirow{4}{*}{\begin{turn}{90}Costs\end{turn}}  &Energy (pJ)&543&633.4&210&6.2\\ 
				\cline{2-6}
				&Current time per decision (pA$\cdot$sec)&452.5&527.8& 311.1& 6.9\\ 
				\cline{2-6}
				&Area (\SI{}{\square\micro\meter})&246,792&133,736&401,300& 2,179\\ 
				\cline{2-6}	
				&Normalized area (\SI{}{\micro\square\meter/\nano\square\meter})& 14.6&7.91&94.9& 1.07\\ 
				\cline{1-6}				
			\end{tabular}
		}		
\end{table}}

\begin{figure}[t]
	\centering
	
	\includegraphics[width=1\columnwidth]{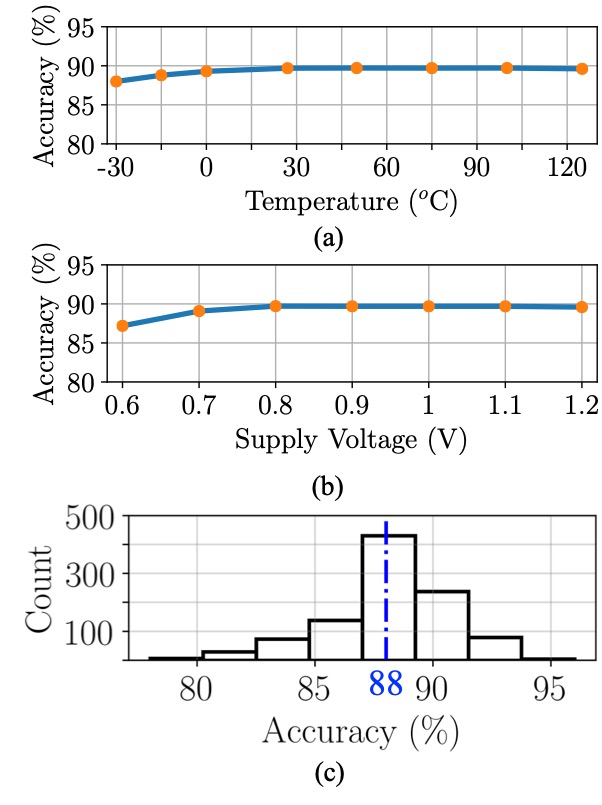} 
	\label{fig:pvt}
	\vspace{-15pt}
	\caption{Classifier performance under PVT variations, (a) under temperature variations at typical process corner, TT, and $V_{DD}$ = \SI{0.9}{\volt}, (b) under voltage variations at room temperature, T = \SI{27}{\celsius}, and at TT, and (c) process and mismatch variations with T = \SI{27}{\celsius} and $V_{DD}$ = 0.9 V.\vspace{-5pt}}
	\label{fig:PVT}
\end{figure}

To evaluate the effect of voltage and temperature variations on the CORE classifier, the supply voltage is varied between 0.6 volts and 1.2 volt and the temperature is varied between \SI{-30}{\celsius} and \SI{125}{\celsius}. The effect of process variations on the classifier performance is evaluated based on a 1K-run Monte-Carlo simulation on a randomly selected 100-observation (10 images per digit) balanced test set with nominal accuracy of 90\%. Note that a 1K-run Monte-Carlo simulation on the whole test set takes $1,000 \times 2.5$ hours on Intel Core i7-7700 CPU. The results of the simulations are shown in Fig. \ref{fig:PVT}. The classifier exhibits no sensitivity within wide range of voltage variations from \SI{0.8}{\volt} to \SI{1.2}{\volt}. Less than 2\% accuracy degradation is observed at low temperatures $\SI{-30}{\celsius}\leq T \leq \SI{0}{\celsius}$. No sensitivity to temperature variations for $\SI{0}{\celsius} \leq T \leq \SI{125}{\celsius}$ is observed. An average of 2\% accuracy degradation is exhibited due to process variations, as extracted from the 1K-run Monte-Carlo simulation. 

Confidence histograms of the correct and incorrect classifications are shown in, respectively, Fig. \ref{fig:confidence_level}(a) and Fig. \ref{fig:confidence_level}(b). With the proposed confidence driven approach, incorrect classifications often exhibit lower confidence as compared with typically confident, correct predictions. The odds of an incorrect classification to be corrected under process variations are therefore high, favorably affecting the resilience of the system to variations. Based on the simulation results, the accuracy is improved for nearly one-third of the Monte-Carlo runs.
\begin{figure}[t]	
	\centering
	\includegraphics[width=1\columnwidth]{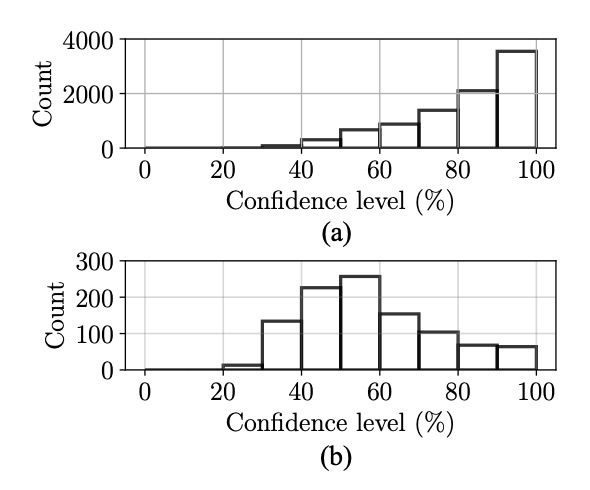} 
	\vspace{-10pt}
	\caption{Confidence level distribution for (a) correct, and (b) incorrect classifications.\vspace{-15pt}} 
	\label{fig:confidence_level}
\end{figure}
\section{CONCLUSION} \label{sec:summary}
Several state-of-the-art mixed-signal classifiers have recently been demonstrated for power efficient classification. 
Accurate classification of multi-dimensional data under the tight power and area constraints is the primary objective in modern on-chip classifiers. 
A novel circuit topology is proposed in this paper for high-COnfidence and high-REsolution (CORE) classification. 
With this topology, body terminals of the MOSFETs are exploited to encode input features, enabling the high-resolution classification. 

To enhance the ML integrity in multi-class classifiers, OVA technique is exploited for efficiently extracting a final decision based on the confidence level of the individual predictors. For a $K$-class classification, $(K-1)/2$ times fewer binary classifiers are required with the OVA approach as compared with the traditional OVO method \cite{zhang2017memory}. 
To further reduce area and power consumption of the OVA-based CORE classifier, a light-weight confidence extractor is designed, generating the final decision based on the confidence level of the individual binary classifiers.
To the best of the authors knowledge, the proposed CORE classifier is the first integrated system to successfully classify MNIST dataset in subthreshold region using a single-MOSFET MAC. Biasing transistors in subthreshold region significantly decreases the leakage and dynamic currents as well as overall load on the sensing lines.

The proposed CORE classifier is designed in SPICE and simulated in \SI{45}{\nano\meter} standard CMOS process. The performance and functionality of the proposed approach is validated with simulation results, exhibiting 90\% classification accuracy with \SI{6.2}{\pico\joule} energy consumption per prediction across the MNIST dataset. Each prediction is finalized within a single clock cycle of \SI{10}{\nano\second}.
The unique topology of CORE classifier supports the ML integrity under a wide range of PVT variations, as well as system scalability across technology nodes.
\bibliographystyle{IEEEtran}

\begin{IEEEbiography}[{\includegraphics[width=1in,height=1.25in,clip,keepaspectratio]{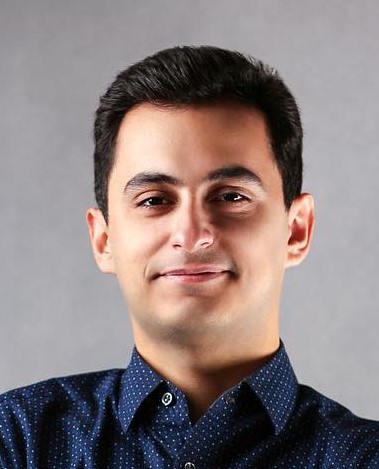}}]{Farid Kenarangi}
(S'18) received the B.Sc. degree in electrical engineering from the University of Tabriz, Tabriz, Iran, in 2015. He is currently pursuing the Ph.D. degree in electrical engineering with The University of Illinois at Chicago, Chicago, IL, USA, under the supervision of Prof. I. Partin-Vaisband. His current research interests include hardware security, machine learning integrated circuits, analog design, and on-chip power delivery and management. He was a recipient of the 2017 University of Illinois at Chicago Chancellor's Graduate Research Award.
\end{IEEEbiography}
\begin{IEEEbiography}[{\includegraphics[width=1in,height=1.25in,clip,keepaspectratio]{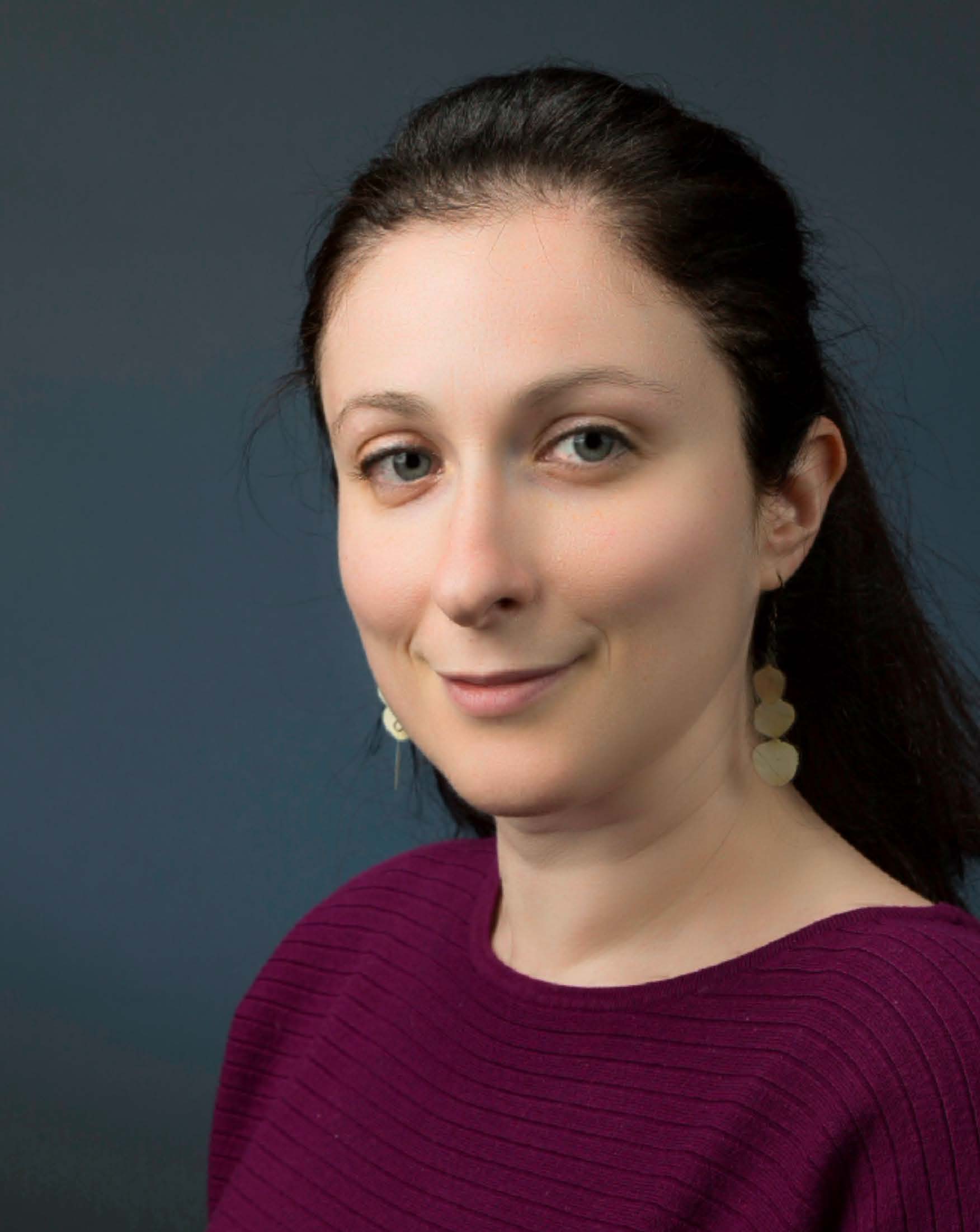}}]{Inna Partin-Vaisband }	
	(S'12--M'15) received the Bachelor of Science degree in computer engineering and the Master of Science degree in electrical engineering from the Technion-Israel Institute of Technology, Haifa, Israel, in, respectively, 2006 and 2009, and the Ph.D. degree in electrical engineering from the University of Rochester, Rochester, NY in 2015. She is currently an Assistant Professor with the Department of Electrical and Computer Engineering at the University of Illinois at Chicago.
	
	Between 2003 and 2009, she held a variety of software and hardware  R\&D positions at Tower Semiconductor Ltd., G-Connect Ltd., and IBM Ltd., all in Israel. Her primary interests lay in the area of high performance integrated circuit design. Her research is currently focused on innovation in the areas of AI hardware and hardware security. Yet another primary focus is on distributed power delivery and locally intelligent power management that facilitates performance scalability in heterogeneous ultra-large scale integrated systems. Special emphasis is placed on developing robust frameworks across levels of design abstraction for complex heterogeneous integrated systems. Dr. P.-Vaisband is an Associate Editor of the \emph{Microelectronics Journal} and has served on the Technical Program and Organization Committees of various conferences.
\end{IEEEbiography}
\end{document}